\documentclass[10pt,journal]{IEEEtran}
\IEEEoverridecommandlockouts
\usepackage{amsmath,bm}
\usepackage[final]{graphicx}
\usepackage{subfigure}
\usepackage{amsthm}
\usepackage{amssymb}
\usepackage{euscript}

\newtheorem{theorem}{Theorem}
\newtheorem{lemma}{Lemma}
\newtheorem{corollary}{Corollary}

\newtheorem{definition}{Definition}
\newtheorem{example}{Example}
\newtheorem*{notation}{Notation}

\newcommand{\defeq}{\overset{\text{def}}{=}}

\begin{document}

\title{On Computing a Function  of Correlated Sources
\thanks{This work was supported in part by a ``Future et Rupture'' grant from
the Institut Telecom, and by an Excellence Chair Grant
 from the French National Research Agency (ACE project).}
\thanks{ M.~Sefidgaran and A.~Tchamkerten are with the
Department of Communications and Electronics, Telecom
ParisTech, 46 Rue Barrault, 75634 Paris Cedex 13, France. Emails: \{sefidgaran,aslan.tchamkerten\}@telecom-paristech.fr.}
\thanks{
Part of this work has been presented at ISIT 2011.
}
}

\author
{
Milad Sefidgaran and Aslan Tchamkerten
}
\maketitle
\begin{abstract}
A receiver wants to compute a function $f$ of two
correlated sources  $X$ and $Y$ and 
side information $Z$. What is the minimum
number of bits that needs to be communicated by each
transmitter?  

In this paper, we derive inner and outer bounds to the rate region of this
problem which coincide in the cases where $f$ is partially invertible 
and where the sources are independent given the side information. 

These rate regions point to an important difference with the single source case. Whereas for the latter it is sufficient to consider independent sets of some suitable characteristic graph, for multiple sources such a restriction is suboptimal and {\it{multisets}} are necessary. 
 \end{abstract}

\section{ Introduction} 
Given two sources $X$ and $Y$ separately observed by
two transmitters, we consider the problem of finding the minimum number of bits
that needs to be sent by each transmitter to a common receiver who has access
to side information $Z$ and wants to compute a given function $f(X,Y,Z)$
with high probability, \textit{i.e.}, with asymptotic zero error probability.\footnote{The results mentioned in this paper are all related to asymptotic zero error probability. Alternatively, (non-asymptotic) zero-error probability has been variously
investigated, {\it{e.g.}}, \cite{Korn73,Orli90,Orli91,Wits76,Yao79,Shay11}.} 

The first result on this problem was obtained by K\"orner and Marton
\cite{KornMar79} who derived the rate region for the case where  $f$ is the sum
modulo two of binary $X$ and $Y$ and where $p(x,y)$ is symmetric (no side
information is available at the receiver). Interestingly, this result came
before Orlitsky and Roche's general result  for the single source case
\cite{OrliRoc01}, which provides a closed form expression on the minimum number
of bits  needed to be transmitted to compute $f(X,Z)$ at the receiver, for
arbitrary $f$ and $p(x,z)$.\footnote{Their result has been generalized for two
round communication \cite{OrliRoc01}, and $ K $ round communication 
\cite{MaIshGup09} in a point-to-point channel.  Also, coding schemes and
converses established in \cite{OrliRoc01}  have been used in  other network
configurations, such as cascade networks~\cite{CuffSuElG09},\cite{visw11}.} However, K\"orner and Marton's  arguments appear to be difficult to generalize to other
functions and probability distributions  (for an extension of \cite{KornMar79}
to sum modulo $ p $ and symmetric distributions see \cite{HanKob87}). Ahlswede and Han \cite{AhlsHan83} proposed an achievable scheme for the sum modulo two problem with an arbitrary probability distribution which is a combination of the  K\"orner-Marton and Slepian-Wolf schemes. The obtained rate region includes, and sometimes strictly, the convex hull of the two schemes. The same scheme has been used in \cite{HuanSko12} to derive an achievable rate region for a certain class of polynomial
functions which is larger than the Slepian-Wolf rate region.  Also, K\"orner-Marton's structural coding scheme has been used to obtain the rate region for certain instances of the problem where the receiver wants to compute some subspace generated by the sources \cite{LaliPraVin11}. 

Except for some specific linear functions and probability distributions, the problem of finding a closed-form expression for the rate region of arbitrary functions and distributions remains in general open. Non closed-form results have been obtained for general functions and distributions by Doshi, Shah, and M\'edard~\cite{DoshShaMed207} who derived
conditions under which a rate pair can be achieved for fixed code
length and error probability. 

A variation of this problem where one of the transmitters observes what the other transmitter sends has been investigated by Ericson and K\"orner \cite{EricKor83}. Because of cooperation,  the rate region of this problem includes the rate region of the problem considered in this paper.

A more general communication setting has been investigated by Nazer and Gastpar
\cite{NazeGas207}, who considered the problem of function computation over a
multiple access channel, thereby introducing potential interference between
transmitters. 

In our problem, we characterize the rate region for a specific function and specific probability distribution. A slightly different problem for the same setting has been considered by Han and Kobayashi \cite{HanKob87}. There, they derived necessary and sufficient conditions for a function, such that for any probability distribution, the rate region of the problem becomes the same as Slepian-Wolf rate region.  Finally, function computation has also been studied in more general networks, such as in
the context of network coding~\cite{AppuMasNik10} and  decentralized decision making and computation \cite{tsit84}.  

In this paper we first provide a general inner bound to the rate region of the function computation problem. Then, we establish an outer bound using results from rate distortion for correlated sources. While this bound is not explicit in general, it implies an explicit outer bound. This latter outer bound and the inner bound are tight for the case where sources are independent given the side information. As a corollary, we recover
the rate region for a single source \cite{OrliRoc01}. Finally, we show that the inner bound characterizes the rate region for partially
invertible functions, \textit{i.e.}, when $X$ or $ Y $ is a function of both $f(X,Y,Z)$
and $Z$.  As a corollary, we recover the Slepian-Wolf
rate region which corresponds to the case where $f(X,Y)=(X,Y)$.

For a single source $X$ and side information $Z$, the minimum number of bits
needed for computing a function $f(X,Z)$ is the solution of an optimization
problem defined over the set of all independent sets with respect to a
characteristic graph defined by $X$, $Z$, and $f$. Indeed, Orlitsky and Roche
showed that, for a single source, allowing for multisets of independent sets
doesn't yield any improvement on achievable rates (see proof of \cite[Theorem
$2$]{OrliRoc01}). By contrast, for multiple sources multisets may
indeed increase the set of achievable rate pairs as we show in an example.

An outline of the paper is as follows. In Section \ref{sec:ProbState} we formally state the
problem and provide some background material and definitions. 
Section \ref{sec:mainResults} contains our results, and Section~\ref{sec:analysis} 
is devoted to the proofs.

\section{Problem Statement and Preliminaries} \label{sec:ProbState} 
\iffalse
\begin{figure} 

\psset{unit=0.04 mm}

\begin{pspicture}(-1500,-1150)
\rput*(0,0){$X$}
\rput*(0,-1000){$Y$}
\rput(980,-500){$Z$}

\psline[arrowsize=5pt]{->}(100,-50)(900,-450)\rput*(560,-160){$R_X$}
\psline[arrowsize=5pt]{->}(100,-950)(900,-550)\rput*(550,-850){$R_Y$}

\rput*(1480,-500){$f(X,Y,Z)$}

\end{pspicture}

\label{fig:setting}
\caption{Setting of the problem}
\end{figure}

\fi
Let $\mathcal{X}$, $\mathcal{Y}$, $\mathcal{Z}$, and $\mathcal{F}$ be finite sets, and $f:\mathcal{X}\times\mathcal{Y}\times
\mathcal{Z}\rightarrow \mathcal{F}$. Let $ (x_i,y_i,z_i),i \geq 1$, be independent instances of
random variables $(X,Y,Z)$ taking values
over $ \mathcal{X} \times \mathcal{Y} \times \mathcal{Z}$ and
distributed according to $p(x,y,z)$.

\begin{definition}[Code]
An $(n,R_X,R_Y)$ code consists of two encoding functions
 \begin{align*}  \varphi_X&:\mathcal{X}^n \rightarrow \{ 1,2,..,2^{nR_X} \}
\nonumber \\ \varphi_Y&:\mathcal{Y}^n \rightarrow \{ 1,2,..,2^{nR_Y} \} \,,
\end{align*} 
and a decoding function
 \begin{align*}  
\psi&: \{ 1,2,..,2^{nR_X} \} \times  \{ 1,2,..,2^{nR_Y} \} \times
\mathcal{Z}^n\rightarrow \mathcal{F}^n\,. \nonumber \end{align*} 
The error probability of a code is defined as
$$P(\psi(\varphi_X(\mathbf{X}),\varphi_Y(\mathbf{Y}),\mathbf{Z})\ne f(\mathbf{X},\mathbf{Y},\mathbf{Z})),$$
where $\mathbf{X}\defeq X_1,\ldots,X_n$ and 
$$f(\mathbf{X},\mathbf{Y},\mathbf{Z})\defeq f(X_1,Y_1,Z_1),...,f(X_n,Y_n,Z_n) \,.$$
\end{definition}
\begin{definition}[Rate Region]
A rate pair $(R_X,R_Y)$ is achievable if, for any $\varepsilon > 0$ and all $n$
large enough, there exists an $(n,R_X,R_Y)$ code whose error probability is no
larger than~$\varepsilon$. The rate region is the closure of the set of achievable 
rate pairs $(R_X,R_Y)$.
\end{definition}
The problem we consider in this paper is to characterize the rate
region for given~$f$ and~$p(x,y,z)$.

We recall the definition of conditional
characteristic graph which plays a key role in coding for
computing. 
 \begin{definition}[Conditional Characteristic Graph  \cite{Korn73,Wits76}] \label{def:CondCharGraph}  Given
$(X,Y)\sim p(x,y)$ and $f(X,Y)$, the conditional
characteristic graph $ G_{X|Y} $  of $X$ given $Y$ is the (undirected)
graph whose vertex set is $\mathcal{X}$ and whose edge set\footnote{We use $ E(G) $ to denote the edge set of a graph $ G $.} $E( G_{X|Y})$ is defined as follows. Two vertices $x_i$ and $x_j$ are connected whenever there exists $y\in \mathcal{Y}$ such that 
\begin{itemize}
\item[i.]
$p(x_i,y) \cdot p(x_j,y) > 0$, 
\item[ii.] $f(x_i,y) \neq f(x_j,y)$.
\end{itemize}
\end{definition}

\begin{notation} Given two random variables $X$ and $V$, where $X$ ranges over
$\cal{X}$ and $V$ over {\emph{subsets}} of $\cal{X}$,\footnote{{\it{I.e.}}, a sample
of $V$ is a subset of $\cal{X}$. An example of a sample of $V$ is $v=\{x_1,x_2\}$, where $x_1,x_2 \in \mathcal{X}$.} we write $ X \in V $
whenever $P(X\in V)=1$.
\end{notation}

Independent sets\footnote{An independent set of a
graph is a subset of its vertices no two of which are
connected. } of a conditional characteristic graph
$G_{X|Y}$ with respect to two random variables $X$ and
$Y$ and a function $f(x,y)$ turns out to be elemental in
coding for computing. In fact, given $Y=y$, the
knowledge of an independent set of $G_{X|Y}$ that includes the
realization $X=x$ suffices to compute $f(x,y)$.

The set of independent sets of a graph $G$
and the set of  maximal independent sets of $G$ are denoted by $\Gamma(G)$ and
$\Gamma^*(G)$, respectively.\footnote{A maximal independent set is an independent set that is not included in any other independent set.}

Given a finite set $\cal{S}$, we use ${\text{M}}({\cal{S}})$ to denote the
collection of all multisets of $\cal{S}$.\footnote{A multiset
of a set $\cal{S}$ is a collection of elements from ${\cal{S}}$ possibly with
repetitions, {\it{e.g.}}, if ${\cal{S}}=\{0,1\}$, then $\{0,1,1\}$ is a multiset.}

\begin{definition} [Conditional Graph Entropy \cite{OrliRoc01}]
\label{def:condGraphEnt} The
conditional entropy of a graph is defined as\footnote{We use the notation $ U-V-W $ whenever random variables 
$ (U,V,W) $ form a Markov chain.} \begin{align*}
H_{G_{X|Y}}(X|Y)&\defeq \min \limits_{\substack{ V -
X- Y\\ X \in V \in \Gamma^*(G_{X|Y})}} I(V;X|Y)\nonumber \\
&=\min \limits_{\substack{ V -
X- Y\\ X \in V \in \text{M}(\Gamma(G_{X|Y}))}} I(V;X|Y).
\end{align*}
\end{definition}

We now extend the definition of conditional
characteristic graph to allow conditioning on variables that take values over
independent sets.

\begin{definition}[Generalized Conditional
Characteristic Graph]  \label{def:GenCondGraph} Given $ (V,X,Y,Z)\sim
p(v,x,y,z)$ and $f(X,Y,Z)$ such that $X \in V \in
\Gamma(G_{X|Y,Z})$,\footnote{By definition
$\Gamma(G_{X|Y,Z})=\Gamma(G_{X|(Y,Z)})$.} define
$$\tilde{f}_X(v,y,z)=f(x,y,z)$$ for $y\in
{\cal{Y}}$, $z\in \cal{Z}$, $x \in v \in
\Gamma(G_{X|Y,Z})$,
and $p(v,x,y,z) > 0$. The generalized conditional
characteristic graph of $Y$ given $V$ and
$Z$, denoted by ${G}_{Y|V,Z} $, is the
conditional characteristic graph of $Y$
given $(V,Z)$ with respect to the marginal distribution 
$p(v,y,z)$ and
$\tilde{f}_X(V,Y,Z)$. \end{definition}

\begin{figure} 
\centering
\subfigure[]{
\includegraphics[scale=.3]{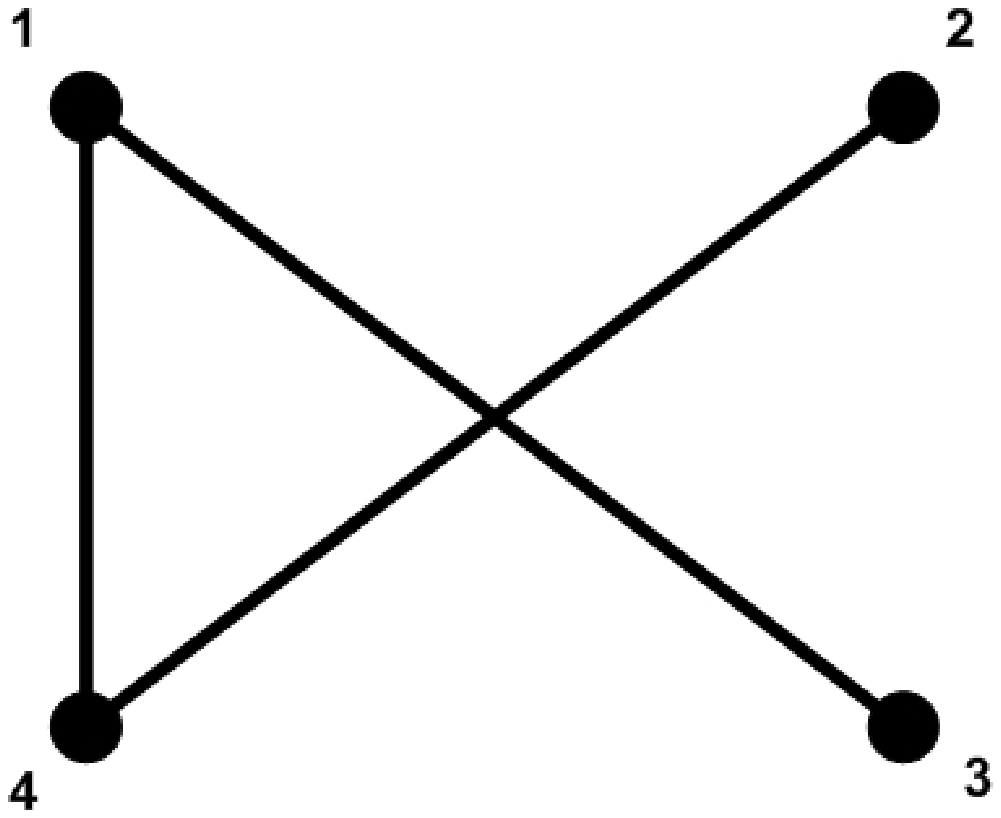}
\label{fig:cond}
}
\subfigure[]{
\includegraphics[scale=.3]{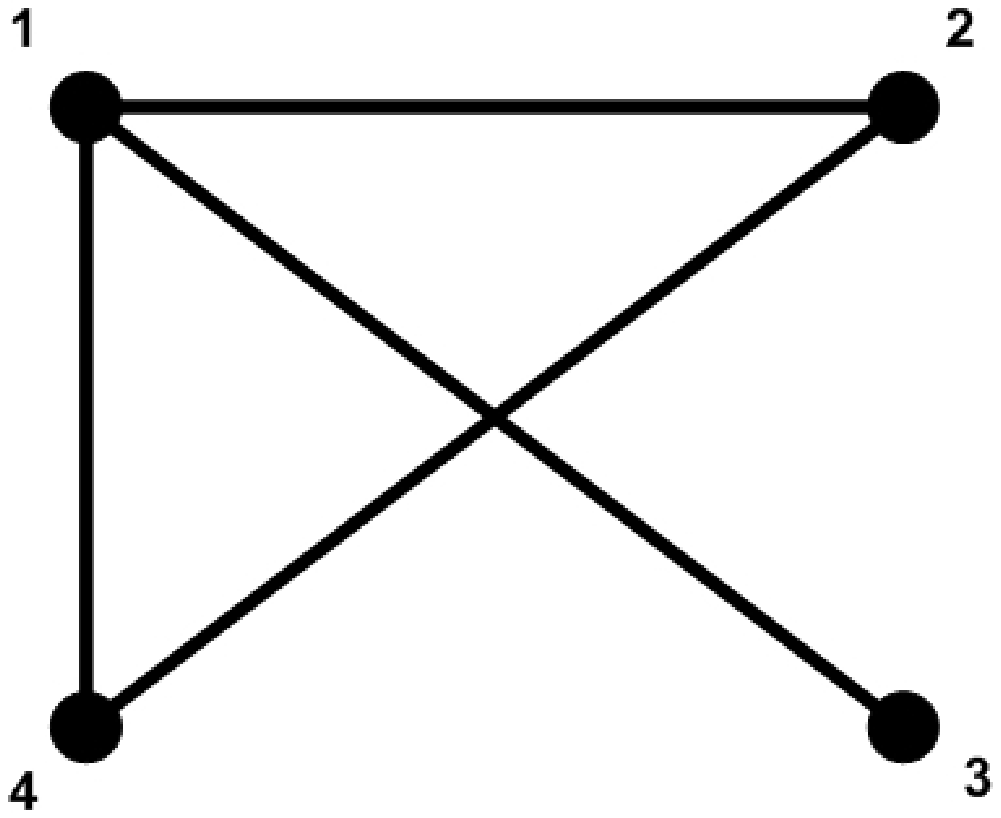}
\label{fig:gencond}
}
\subfigure[]{
\includegraphics[scale=.3]{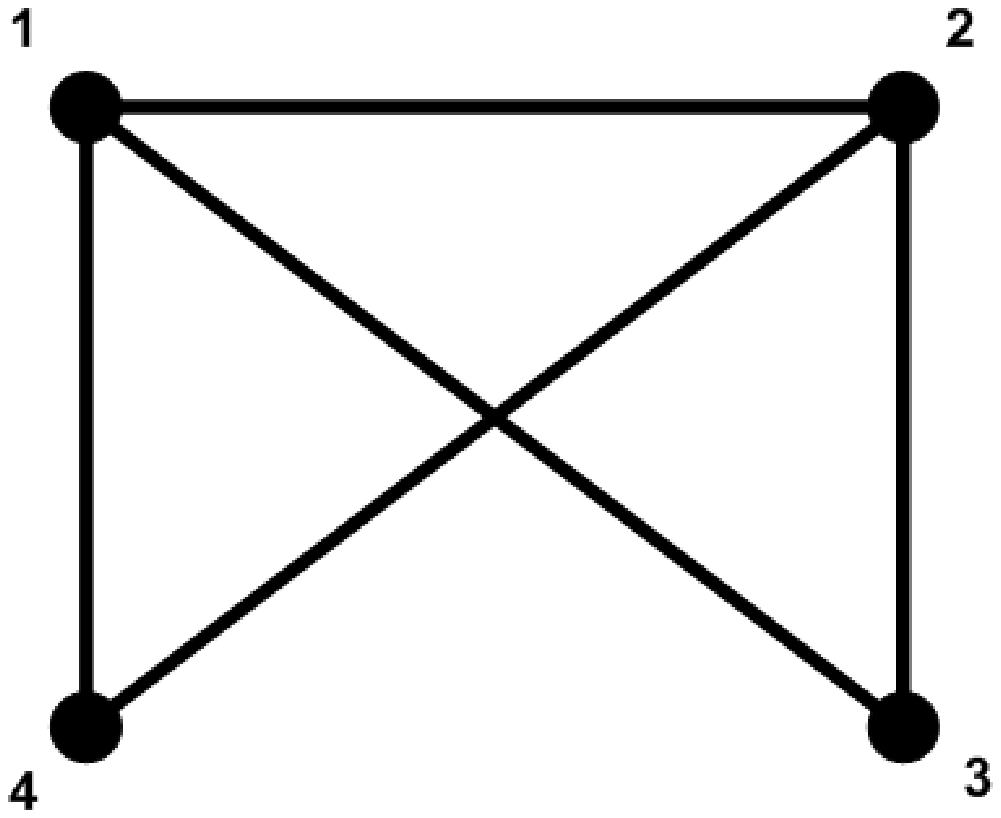}
\label{fig:gencond2}
}

\label{fig:examplecond}
\caption{\subref{fig:cond} $ G_{X|Y} $ and $ G_{Y|X} $, 
\subref{fig:gencond} and \subref{fig:gencond2} $ G_{X|W} $.}
\end{figure}

\begin{example}  \label{ex:bigger}
Let $ X $ and $ Y $  be random 
variables defined over the alphabets $ \mathcal{X}$ and 
$\mathcal{Y}$, respectively, with  $$ \mathcal{X}=\mathcal{Y}=\{1,2,3,4\} \,.$$
Further, suppose that  $ p(X=Y)=0 $ and that $(X,Y)$ take on values uniformly over the pairs $ (i,j) \in \mathcal{X} \times 
\mathcal{Y}$ with $ i \neq j $. The receiver wants 
to decide whether $ X >Y $ or $ Y >X $, {\it{i.e.}}, it wants to compute\begin{gather}
\raisetag{-10pt}
f(x,y) = \begin{cases}
0 \mbox{  if } x<y,\\
1  \mbox{ if } x>y. 
\end{cases}  \nonumber
\end{gather}

Fig. \ref{fig:cond} depicts $ G_{X|Y}$ which is 
equal to $ G_{Y|X} $ by symmetry. Hence we have
\begin{align*}
 \Gamma(G_{X|Y})&=\Gamma(G_{Y|X})\\
 &=\{ \{1\},\{2\},\{3\},\{4\},\{1,2\},\{2,3\},\{3,4\} \},
\end{align*}
and
$$ \Gamma^*(G_{X|Y})=\Gamma^*(G_{Y|X})=\{ \{1,2\},\{2,3\},\{3,4\} \}. $$
An example of a random variable $V$ that satisfies 
\begin{align}\label{eq:YW}
X \in V \in  \Gamma(G_{X|Y})
\end{align}
is one whose support set is
$$ \mathcal{V}=\{ \{1\},\{2\},\{3\},\{4\},\{1,2\} \}. $$ 
For such a $V$, the generalized conditional characteristic graph $ G_{Y|V} $ 
 is depicted in Fig.~\ref{fig:gencond} and we have 
$$ \Gamma(G_{Y|V})=\{ \{1\},\{2\},\{3\},
\{4\},\{2,3\},\{3,4\} \}. $$
Another $V$ that satisfies \eqref{eq:YW} is one whose
support set is 
$$ \mathcal{V}=\{\{2\},\{4\},\{1,2\},\{2,3\} \}. $$
For such a $V$, the generalized conditional characteristic graph $ G_{Y|V} $ 
is depicted in Fig. \ref{fig:gencond2} and we have 
$$ \Gamma(G_{Y|V})=\{ \{1\},\{2\},\{3\},
\{4\},\{3,4\} \}. $$
\end{example}
Note that $$ E(G_{Y|X,Z}) \subseteq E(G_{Y|V,Z}) $$ 
whenever 
$$X\in V\in \Gamma(G_{X|Y,Z}).$$ 
The following lemma, proved in Section~\ref{sec:analysis},
 provides sufficient conditions under
which  $$E(G_{Y|X,Z})=E(G_{Y|V,Z}).$$
\begin{lemma} \label{lem:samecondgraph}
Given $ (V,X,Y,Z) \sim p(v,x,y,z)$ and $f(X,Y,Z) $, we have
 $$ G_{Y|V,Z}=G_{Y|X,Z} $$
for all $V$ such that $X\in V\in
\Gamma(G_{X|Y,Z})$
 in each of the following cases:
\begin{itemize}
\item[a.]
$ p(x,y,z) >0 $ for all $ (x,y,z) \in \mathcal{X} \times \mathcal{Y} \times \mathcal{Z} $;
\item[b.]
$G_{X|Y,Z}$ is a complete graph or, equivalently, 
$ \Gamma(G_{X|Y,Z}) $ consists only of singletons;
\item[c.]
$ X $ and $ Y $  are independent given $Z$.
\end{itemize}
\end{lemma}
Notice that if $G(X|Y,Z)$ is a complete graph, by knowing
 $Y$ and $Z$ the function $f(X,Y,Z)$
can be computed only if also $X$
is known exactly.

\section{Results} \label{sec:mainResults}
Our results are often stated in terms of certain random variables $V$ and $W$
which can usefully be interpreted as the messages sent by transmitter-X and transmitter-Y, respectively. This interpretation is 
consistent with the proofs of the results.

\subsection{Inner Bound}
Theorem~\ref{achiev} provides a general
inner bound to the rate region: 
\begin{theorem}[Inner bound] \label{achiev} $ (R_X,R_Y) $ is achievable whenever
  \begin{align} R_X &\geq I(V;X|W,Z) \nonumber \\ R_Y &\geq
I(Y;W|V,Z) \nonumber \\ R_X+R_Y &\geq I(V;X|Z)+I(Y;W|V,Z),\nonumber 
\end{align} for some $V \in \mathcal{V}$ and $W \in \mathcal{W}$ that satisfy the Markov chain constraints
\begin{align}
V - X - (Y,W,Z)\nonumber \\
(V,X,Z) - Y - W, \label{eq:achieveMarkov}
\end{align}
and either 
\begin{align}
X \in V \in \text{M}(\Gamma(G_{X|Y,Z})) \nonumber \\
Y \in W \in \text{M}(\Gamma(G_{Y|V,Z})), \label{eq:achieveSet1}
\end{align}
or, equivalently, 
\begin{align}
Y \in W \in \text{M}(\Gamma(G_{Y|X,Z})) \nonumber \\
X \in V \in \text{M}(\Gamma(G_{X|W,Z})).  \label{eq:achieveSet2}
\end{align}
Moreover, we have the following cardinality bounds on the range of $V$ and $W$:
\begin{align*}
|\mathcal{V}|\hspace{.1 cm}&\leq |\mathcal{X}|+1\\
|\mathcal{W}|&\leq |\mathcal{Y}|+1.
\end{align*}
When there is no side information at the decoder, {\it{i.e.}}, when $ Z$ is a
constant, the two Markov chain constraints are equivalent to the single long Markov chain $$
V-X-Y-W\,, $$ which imply that the above sum rate inequality
becomes $$R_X+R_Y\geq
I(X,Y;V,W)\,.$$ 
\end{theorem}
The last part of the theorem is immediate.

Note that in the above theorem, $V$ and $W$ are not
restricted to take values over maximal independent
sets. By contrast with the single source case where the
restriction to 
maximal independent induces no loss of optimality---see
Definition~\ref{def:condGraphEnt} where $V$ may be restricted to range
over $\Gamma^*(G_{X|Y})$---for two sources the restriction to maximal independent sets may
indeed induce a loss of optimality. This will be illustrated
in Example~\ref{ex:partiallyinv} of
Section~\ref{rregions} which considers a setting where
Theorem~\ref{achiev} is tight and characterizes the
rate region.

Theorem~\ref{achiev} does not, in general, give the
rate region. An example of this is the sum modulo $2$ of binary $X$ and $Y$ (no
side information) with symmetric
distribution as considered by K\"orner and
Marton~\cite{KornMar79}:
\begin{example}\label{exKM}
Let $f(X,Y)$ be the sum modulo $2$ of binary $X$ and
$Y$ with joint
distribution  
$$p(x,y)= \left[ \begin{array}{cc} \frac{p}{2}& \frac{1-p}{2} \\ \frac{1-p}{2} & \frac{p}{2} \end{array}
\right].$$
Assuming $p\in (0,1)$, $ \Gamma(G_{X|Y}) $ and $ \Gamma(G_{Y|X}) $  both
consists of singletons. This implies that the achievable region given by
Theorem~\ref{achiev} reduces to
 \begin{align} 
 R_X &\geq H(X|W) \nonumber \\ 
 R_Y &\geq H(Y|V) \nonumber \\ 
 R_X+R_Y &\geq H(X)+H(Y|V),\label{eq:iter}
\end{align}
since $$H(X|V)=H(Y|W)=0$$
for all $(V,X,Y,W)$ that satisfy
$$X \in V \in \text{M}(\Gamma(G_{X|Y}))$$ 
$$Y \in W \in \text{M}(\Gamma(G_{Y|V})).$$
Note that since $ \Gamma(G_{Y|V}) $ (which is equal to $ \Gamma(G_{Y|X}) $ 
according to Claim a. of Lemma \ref{lem:samecondgraph}) consists of singletons, we have
\begin{align}
H(X|W)=H(X|Y,W)\leq H(X|Y).  \label{eq:xwless}
\end{align}
Furthermore, because of the Markov chain constraint
$$(V,X)-Y-W,$$
we have 
\begin{align} 
H(X|W)\geq H(X|Y) \label{eq:xwmore}
\end{align}
by the data processing inequality. Hence, \eqref{eq:xwless} and \eqref{eq:xwmore} yield
$$H(X|W)= H(X|Y), $$
and, from the same argument we get
$$H(Y|V)= H(Y|X).$$

Inequalities \eqref{eq:iter} thus 
become 
 \begin{align} 
 R_X &\geq H(X|Y) \nonumber \\ 
 R_Y &\geq H(Y|X) \nonumber \\ 
 R_X+R_Y &\geq H(X,Y)\label{eq:iter2}
\end{align}
which corresponds to the Slepian-Wolf rate region. This region isn't maximal
since the maximal rate region is given by the set of rate pairs 
that satisfy the only two constraints
\begin{align*} 
 R_X &\geq H(X|Y)  \\
 R_Y &\geq H(Y|X) 
\end{align*}
as shown by K\"orner and Marton~\cite{KornMar79}.
\end{example}

\subsection{Outer Bounds}
We now provide a rate region outer bound which is
derived using results from rate distortion for correlated sources \cite{Tung78}: 
\begin{theorem}[Outer Bound I] \label{th:OuterBound1}
 If  $ (R_X,R_Y) $ is achievable, then
  \begin{align} R_X &\geq I(X,Y;V|W,Z) \nonumber \\ R_Y &\geq
I(X,Y;W|V,Z) \nonumber \\ R_X+R_Y &\geq I(X,Y;V,W|Z),\nonumber 
\end{align} 
for some random variables $(V,W)$ 
that satisfy $H(f(X,Y,Z)|V,W,Z)=0$ and   Markov chain constraints
$$V - X - (Y,Z)$$
$$(X,Z) - Y - W$$
$$(V,W)-(X,Y)-Z.$$
\end{theorem}
\iffalse
From Lemma \ref{lem:jointCompressionLong} in the
Appendix A, one can show that
had the above first two Markov chain constraints been  $$V - X - (Y,W,Z)$$ $$(V,X,Z)
- Y - W,$$ the outer bound given by Theorem~\ref{th:OuterBound1} would be
equal to the inner bound given by Theorem~\ref{achiev}. However, 
these hypothetical Markov chains don't hold in general
since the inner bound is not tight in general as we
show in Example~\ref{exKM}. 
\fi
Although Theorem~\ref{th:OuterBound1} doesn't provide an explicit outer bound---it is implicitly characterized by the random
variables $(V,W)$ that should (in part) satisfy $H(f(X,Y,Z)|V,W,Z)=0$---this theorem implies the following explicit outer bound which can alternatively be derived 
from \cite[Theorem $1$]{OrliRoc01}:
\begin{corollary}[General Outer Bound 2] \label{th:OuterBound2}
If $ (R_X,R_Y) $ is achievable then
  \begin{align} R_X &\geq H_{G_{X|Y,Z}}(X|Y,Z) \nonumber \\ R_Y &\geq
H_{G_{Y|X,Z}}(Y|X,Z) \nonumber \\ R_X+R_Y &\geq H_{G_{X,Y|Z}}(X,Y|Z) \nonumber 
\end{align}
\end{corollary}
\subsection{Rate Regions}\label{rregions}
The inner and outer bounds given by
Theorem~\ref{achiev} and
Corollary~\ref{th:OuterBound2} are
tight for independent sources, hence also for the single
source computation problem\footnote{A single source 
can be seen as two sources with one of them being constant.} for which we recover 
\cite[Theorem~$1$]{OrliRoc01}. When the sources are conditionally independent given the side information, the rate region is the solution of two separate point-to-point problems. This is analogous to a result of Gastpar \cite{Gast04} which says that under the independence condition the rate-distortion region for correlated sources is the solution of two separate point-to-point Wyner-Ziv problems.

\begin{theorem} [Rate Region - Independent Sources]\label{cor:independent}
If $ X $ and $ Y $ are independent 
given $ Z $, the rate region is the closure of rate pairs
$ (R_X,R_Y) $ such that
\begin{align}
R_X \geq H_{G_{X|Y,Z}}(X|Y,Z) \nonumber \\
R_Y \geq H_{G_{Y|X,Z}}(Y|X,Z). \nonumber 
\end{align}
Hence, if $Y$ is constant, $R_X$ is achievable if and only if $R_X\geq H_{G_{X|Z}}(X|Z) $. 
\end{theorem}
\begin{example}\label{ex:single}
Let $ Z \in \{ 1,2,3 \} $, let $ U $ and $ V $
be independent uniform random variables over
$ \{-1,0,1\} $ and $ \{0,1,2\} $, respectively,  and let
$ X=Z+U $ and $ Y=Z+V $. The receiver wants to compute the function 
$ f(X,Y) $ defined as
\begin{gather}
\raisetag{-10pt}
f(x,y) = \begin{cases}
0 \mbox{  if } x \neq y,\\
1  \mbox{ if } x=y. 
\end{cases} \nonumber
\end{gather}
Since $ X $ and $ Y $ are independent
given $ Z $, the rate region is given by Theorem~\ref{cor:independent}. 
It can be checked that
$$ \Gamma^*(G_{X|Y,Z})=\{ \{0,2\}, \{ 0,3\} ,\{0,1,4\}\} $$
$$ \Gamma^*(G_{Y|X,Z})=\{ \{2,5\}, \{ 3,5\} ,\{1,4,5\}\}\,,$$
and a numerical evaluation of conditional graph entropy gives 
$$H(G_{X|Y,Z})=H(G_{Y|X,Z}) \simeq 1.28\,. $$ Hence the rate region is given by the
set of rate pairs satisfying 
\begin{align}
R_X & \gtrsim 1.28 \nonumber \\
R_Y & \gtrsim 1.28. \nonumber 
\end{align}
\end{example}

The following theorem gives the rate region when the function is
partially invertible with respect to $X$ (with respect to $Y$, respectively),
{\it{i.e.}}, when $X$ ($Y$, respectively) is a deterministic function of both
$f(X,Y,Z)$ and $Z$.\footnote{A similar definition is given in \cite{EricKor83}, in a way that $f(X,Y)$ is partially invertible if $H(
X|f(X,Y),Y)=0$.} 

\begin{theorem}[Rate Region - Partially Invertible Function] \label{CapacityOnecomplete} If
$f$ is partially invertible with respect to $X$, then the rate region is the
closure of rate pairs $ (R_X,R_Y) $ such that \begin{align} R_X &\geq H(X|W,Z)
\nonumber \\ R_Y &\geq I(Y;W|X,Z) \nonumber \\ R_X+R_Y &\geq
H(X|Z)+I(Y;W|X,Z), \nonumber  \end{align} for some $W \in \mathcal{W}$ that satisfies 
$$(X,Z) - Y - W$$ $$Y \in W \in  \text{M}(\Gamma(G_{Y|X,Z})),$$ with the following cardinality bound
$$ |\mathcal{W}| \leq |\mathcal{Y}|+1.$$
\end{theorem}

When $f$ is invertible, $(X,Y)$ is a function of both $f(X,Y,Z )$ and $Z$,
and Theorem~\ref{CapacityOnecomplete} reduces to the Slepian-Wolf rate region~\cite{SlepWol73}.

\iffalse
\begin{corollary}[Rate Region - Invertible Function]\label{cor:invertible}
If $ f $ is invertible, then the rate region
 is the closure of rate pairs $ (R_X,R_Y) $ such
that  \begin{align}
 R_X &\geq H(X|Y,Z), \nonumber \\
 R_Y &\geq H(Y|X,Z), \nonumber \\
 R_X+R_Y &\geq H(X,Y|Z). \nonumber 
\end{align}
\end{corollary}

The above result is trivial, since when the function is invertible, recovering the function is equivalent to recovering the sources. In the Appendix, it has been shown that this result could be derived from Theorem~\ref{CapacityOnecomplete}. 

\begin{example} Let  
$ \mathcal{X}=\mathcal{Y} = \{ 1,2,3,5,7 \} $, 
$\mathcal{Z}=\{ 1,2,3\}$, let $p(x,y,z)$ be such that $$ p(x,y,z) \cdot
p(y,x,z)=0, $$ for any $x,y,z$ with $x\ne y$, and let
$$ f(x,y,z)=x \cdot y \cdot z \,. $$ Since $ f(X,Y,Z) $ 
is invertible, the rate region is the 
Slepian-Wolf rate region given by Corollary~\ref{cor:invertible}.
\end{example}
\fi

\begin{example}
\label{ex:partiallyinv}

\begin{figure}  \centerline{\includegraphics[width=10cm]{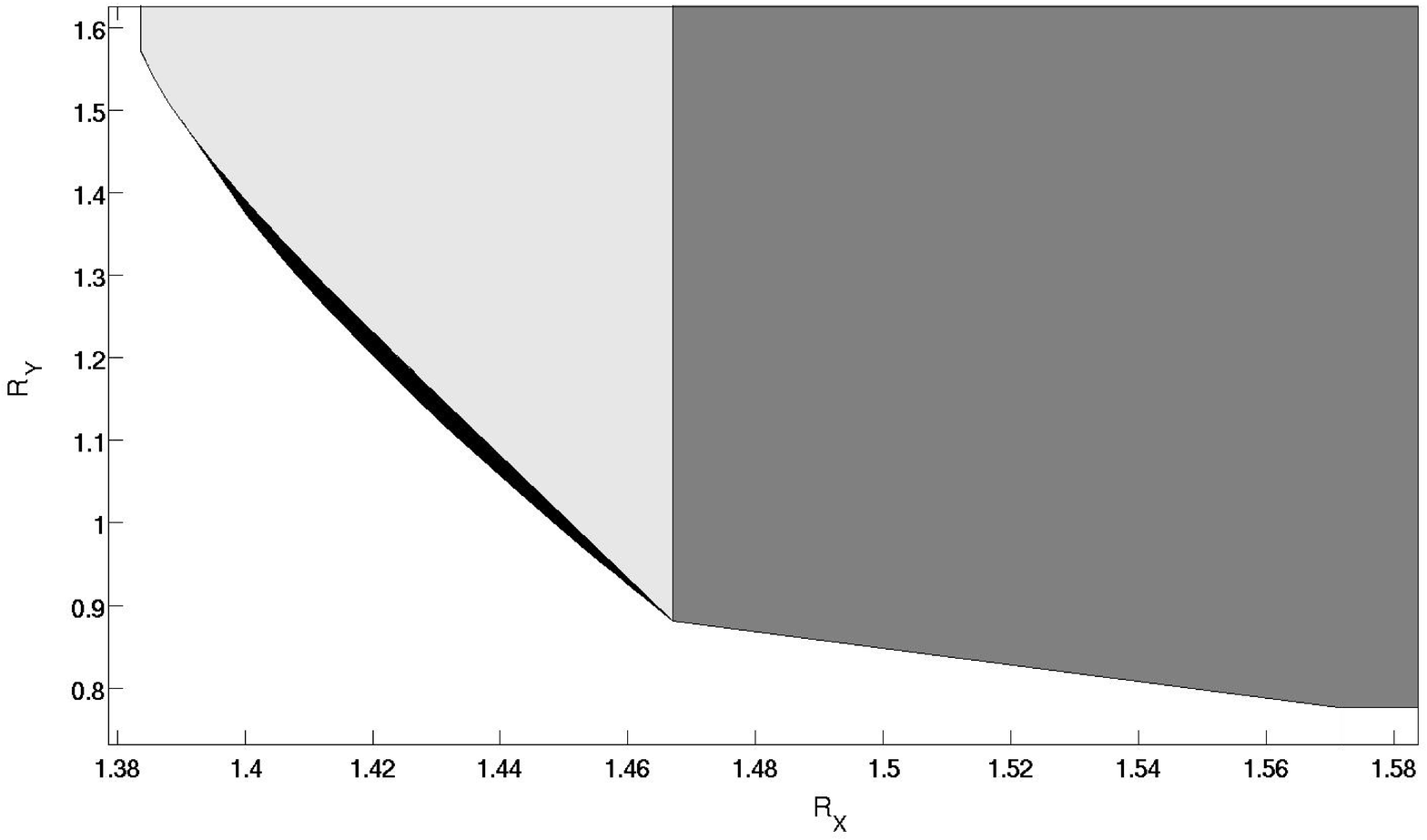}} \caption{\label{fig:achiev}
Example of a rate region for a partially invertible function.} \end{figure}
 Consider the situation with
no side information given by $f(x,y)= (-1)^y \cdot x $,
with $\mathcal{X}=\mathcal{Y}=\{0,1,2\}$,  and
$$p(x,y)= \left[ \begin{array}{ccc} .21& .03& .12 \\ .06& .15& .16 \\  .03& .12& .12 \end{array}
\right].$$  Since $f(X,Y)$ is partially invertible with respect to $X$, we can
use Theorem~\ref{CapacityOnecomplete} to numerically evaluate the rate region. 
The obtained region is given by the union of the three shaded areas in
Fig.~\ref{fig:achiev}. These
areas are discussed later, after Example~\ref{ex:single}.

To numerically evaluate the rate region, we would need to consider the set of all conditional
distributions  $p(w|y)$, $y\in {\cal{Y}}$, $w\in \text{M}(\Gamma(G_{Y|X})) $.
Since $|\mathcal{W}|\leq 4$, 
$\text{M}(\Gamma(G_{Y|X}))$ consists of multisets 
of  
$$\Gamma (G_{Y|X})=\{ \{0 \},\{1 \},\{2\}, \{ 0,2\}\} $$
whose cardinalities are bounded by $4$.

However, as we now show, 
among all possible $ 4^4=256 $ multisets with cardinality at most $4$,
 considering just the multiset 
$ \{\{1\},\{0,2\},\{0,2\},\{0,2\}\} $ gives the rate region. 

Consider a multiset with cardinality at most $4$.
\begin{itemize}
\item[1.] If the multiset does not contain
 $ \{1\} $, then the condition $ \sum \limits_{w \in \mathcal{W}} p(w|Y=1)=1 $, 
hence the condition $ Y \in W $, cannot be satisfied. Therefore this multiset is
not admissible, and we can ignore it.  
\item[2.]
 If the multiset contains two samples $ w_1=\{1\} $ and $ w_2=\{1\} $  with
conditional probabilities $ p(w_1|Y=1) $ and $ p(w_2|Y=1) $, respectively, 
replacing them by one sample $ w=\{1\} $ whose conditional probability is
$p(w|Y=1)= p(w_1|Y=1)+p(w_2|Y=1) $, gives  the same terms $ H(X|W) $ and $
I(Y;W|X) $, hence  the same rate pairs. Therefore, without loss of optimality
we can consider only multisets which contain a unique sample of $ \{1\} $.

\item[3.] If the multiset contains a sample $ w_1=\{0\} $ with arbitrary
conditional  probability $ p(w_1|Y=0) $, replacing it with sample $ w_2=\{0,2\}
$ whose conditional probabilities are $ p(w_2|Y=0)=p(w_1|Y=0) $ and $
p(w_2|Y=2)=0 $ gives the same rate pairs. (The same argument holds
for a sample $ w_1=\{2\} $).  

From $1.$, $2.$, and $3.$, multisets with one sample of
$\{1\} $ and multiple copies of $\{0,2\} $ gives the rate region.

\item[4.] If the multiset has cardinality $ k <4 $,  adding $ 4-k $ samples 
$ \{0,2\} $ with zero conditional probabilities, gives the same rate pairs.
\end{itemize}

It follows that the rate region can be obtained by considering the unique multiset $$
\{w_1=\{1\},w_2=\{0,2\},w_3=\{0,2\},w_4=\{0,2\}\} $$ and by
optimizing over the conditional probabilities $\{p(w|y)\}$ that satisfy
$$p(w_1|Y=1)=1$$
$$p(w_1|Y=j)=0,j \in \{0,2\}$$
$$ \sum \limits_{i=2}^4 p(w_i|Y=0)=1$$
$$ \sum \limits_{i=2}^4 p(w_i|Y=2)=1$$
$$p(w_i|Y=1)=0,i \in \{2,3,4\}.$$
Notice that this optimization has only four degrees of freedom.

Fig.~\ref{fig:achiev} shows the achievable 
rate region in Theorem~\ref{achiev} when
restricting $V$ and $W$ to be over maximally independent sets (gray area), all
independent sets (gray and light gray areas), and multisets of independent sets
(union of gray, light gray, and black areas). The latter area corresponds 
to the rate region by Theorem~\ref{CapacityOnecomplete}. Denoting these areas by 
${\cal{R}}(\Gamma^*)$, ${\cal{R}}(\Gamma)$, and ${\cal{R}}(\text{M}(\Gamma))$,\footnote{With $|\text{M}(\Gamma)|\leq 5$.} respectively, we thus numerically get the strict sets inclusions
$${\cal{R}}(\Gamma^*)\subset {\cal{R}}(\Gamma)\subset
{\cal{R}}(\text{M}(\Gamma)).$$

Larger independent sets for $X$
allow to reduce $R_X$. However, such sets may have less correlation with $Y$ and
$Z$, and so may require to increase $R_Y$. By contrast, for the single source case,
since only $R_X$ needs to be minimized it is
optimal to choose maximal independent sets. 
\iffalse but here
choosing a bigger independent set may be non-optimal,
and we have a trade-off between $R_X$ and $R_Y$.
Moreover, this holds also for two same independent sets
with different labelings(different conditional
probabilities $p(v|x)$) , {\it{i.e.}} replacing two
independent sets by one of them may lead into greater
(or smaller) $R_X$ and smaller (or greater) $R_Y$,
which means that it may be non-optimal to replace a
multiset by a set.
\fi
Numerical evidence suggests that the small difference between ${\cal{R}}(\Gamma)$ and
${\cal{R}}(\text{M}(\Gamma))$ is unrelated to the specificity of the
probability distribution $p(x,y)$ in the example ({\it{i.e.}}, by choosing other
distributions the difference between ${\cal{R}}(\Gamma)$ and
${\cal{R}}(\text{M}(\Gamma))$ remains small).

\end{example}

\section{Analysis} \label{sec:analysis}

\begin{IEEEproof} [Proof of Lemma \ref{lem:samecondgraph}]
Suppose $ X \in V \in \Gamma(G_{X|Y,Z}) $. For all claims, we show that $E(G_{Y|V,Z}) \subseteq E(G_{Y|X,Z})$, {\it{i.e.}}, if two nodes are connected in $ G_{Y|V,Z} $, then they are also connected in $ G_{Y|X,Z} $. The opposite direction, $E(G_{Y|X,Z}) \subseteq E(G_{Y|V,Z}) $, follows from the definition of generalized conditional characteristic graph. 

Suppose nodes $ y_1 $  and $ y_2  $ are connected 
in  $ G_{Y|V,Z} $. This means that there exist 
$ v \in \mathcal{V} $, $ x_1, x_2 \in v $  and 
$ z \in \mathcal{Z} $ such that 
$$ p(x_1,y_1,z) \cdot p(x_2,y_2,z)>0, $$ and 
$$ f(x_1,y_1,z)\neq f(x_2,y_2,z). $$ If $ x_1=x_2 $, 
then $ y_1 $  and $ y_2  $ are also connected in  
$ G_{Y|X,Z} $ according to the definition of conditional
characteristic graph. We now assume 
$ x_1 \neq x_2 $ and prove Claims a., b., and c.

\begin{itemize} 
\item[a.] Since all probabilities are positive 
we have $ p(x_1,y_2,z)>0 $, hence 
$$ p(x_1,y_1,z)\cdot p(x_1,y_2,z)>0, $$ and $ x_1,x_2 \in v 
\in \Gamma(G_{X|Y,Z}) $ yields $$ f(x_1,y_2,z)=f(x_2,y_2,z) \neq f(x_1,y_1,z), $$ 
which implies that $ y_1 $  and $ y_2  $ are also connected in  $ G_{Y|X,Z} $.
\item[b.] $ \Gamma(G_{X|Y,Z}) $ consists of singletons, 
so $ x_1,x_2 \in v 
\in \Gamma(G_{X|Y,Z}) $ yields
$ x_1=x_2 $, and thus $ y_1 $  and $ y_2  $ 
are also connected in  $ G_{Y|X,Z} $ as we showed above.
\item[c.] From the independence of $ X $ and $ Y $ 
given $ Z $ we have $$ p(x,y,z)=p(z) \cdot p(x|z)\cdot p(y|z).$$ 
Hence, since $$ p(x_1,y_1,z) \cdot p(x_2,y_2,z)>0, $$ 
we have 
 $$ p(z) \cdot p(x_1|z) \cdot p(y_2|z) >0, $$  
 {\it{i.e.},} $ p(x_1,y_2,z)>0 $. The
rest of the proof is the same as Claim a..
\end{itemize}
\end{IEEEproof}

\begin{IEEEproof}[Proof of Theorem \ref{achiev}] We consider a coding scheme similar to the Berger-Tung rate distortion coding
scheme \cite{Tung78} with the only difference that here we use jointly robust typicality \cite{OrliRoc01} 
in place of strong typicality. 
Recall that $ (\bm{v},\mathbf{x}) $ are jointly $ \delta $-robust typical \cite{OrliRoc01},
if
$$|\tilde{p}_{\mathbf{v},\mathbf{x}}(v,x)-p(v,x)|\leq \delta \cdot p(v,x)$$
for all $ (v,x) \in \mathcal{V} \times \mathcal{X} $, where 
$$\tilde{p}_{\mathbf{v},\mathbf{x}}(v,x) \defeq \frac{|\{i:(v_i,x_i)=(v,x)\}|}{n}.$$
Note that if $ (\mathbf{v},\mathbf{x}) $ 
are jointly robust typical, then $\forall i, p({v}_i,{x}_i)>0 $, and if $V$ takes values over subsets of $\mathcal{X}$, it means that $ \forall i, x_i \in v_i$. This fact ensures the asymptotic zero block error probability in decoding the random variables $V$ and $W$. Moreover, having decoded them reliably and since $V$ and $W$ satisfy Markov chains \eqref{eq:achieveMarkov}, asymptotic zero block error probability in decoding the function $f$ follows.

For the rest of the proof, first we show the equivalence of the conditions \eqref{eq:achieveSet1} and \eqref{eq:achieveSet2}, and then we establish the cardinality bounds.

For equivalence, we prove one direction, the proof for the other direction is analogues. Suppose that \eqref{eq:achieveSet1} holds, \textit{i.e.}
\begin{align*}
X \in V \in \text{M}(\Gamma(G_{X|Y,Z})) \\
Y \in W \in \text{M}(\Gamma(G_{Y|V,Z})). \nonumber 
\end{align*} 
To prove that $  W \in \text{M}(\Gamma(G_{Y|X,Z})) $,  
we show that for any $ w \in \mathcal{W}$,
$ y_1, 
y_2 \in w $, $ x \in \mathcal{X} $, and $ z 
\in \mathcal{Z} $ such that 
$$ p(x,y_1,z) \cdot p(x,y_2,z)>0, $$ 
we have 
$$ f(x,y_1,z)=f(x,y_2,z). $$ 
Since $ P(X \in V)=1 $, 
there exists $ v \in V $ such that $ p(v|x) >0 $, 
hence, by the definition of generalized conditional 
characteristic graph $ G_{Y|V,Z} $, we have
$$ f(x,y_1,z)=\tilde{f}_X(v,y_1,z)=\tilde{f}_X(v,y_2,z)=f(x,y_2,z). $$
To prove that $ V \in \text{M}(\Gamma(G_{X|W,Z})) $, 
note that for any $ w \in \mathcal{W}$, 
$ y_1, y_2 \in w $, $ v \in \mathcal{V} $, 
$ x_1,x_2 \in \mathcal{X} $, and 
$ z \in \mathcal{Z} $ such that 
$ p(x,y_1,z) \cdot p(x,y_2,z)>0 $,

\begin{itemize}
\item[i)] if $ y_1 = y_2=y $, then $ f(x_1,y,z)=f(x_2,y,z)$, since $V \in \text{M}(\Gamma(G_{X|Y,Z}))$.
\item[ii)] if $ y_1 \neq y_2 $, then   
$$ f(x_1,y_1,z)=\tilde{f}_X(v,y_1,z)=\tilde{f}_X(v,y_2,z)=f(x_2,y_2,z)$$ 
since  $W \in \text{M}(\Gamma(G_{Y|V,Z}))$. 
\end{itemize}
Hence $V \in \text{M}(\Gamma(G_{X|W,Z}))$ for both 
cases i) and ii).

Now, we prove the cardinality bounds using the method described in \cite[Appendix C]{ElgaKim12}. We show that
$$|\mathcal{V}| \leq |\mathcal{X}|+1.$$
The proof for
$$|\mathcal{W}| \leq |\mathcal{Y}|+1$$
is analogous.

Random variables $(V,X,Y,W,Z)$ as in the Theorem have joint distribution of the form
$$p_{V,X,Y,W,Z}(v,x,y,w,z)=p_{V}(v)p_{X|V}(x|v)p_{Y,W,Z|X}(y,w,z|x).$$
In the following, we show that it is possible to replace $V \sim p_V(\cdot)$ and $p_{X|V}(\cdot|\cdot)$ with
$V' \sim p_{V'}(\cdot)$ and  $p_{X|V'}(\cdot|\cdot)$ such that 
\begin{itemize}
\item $p_X$ and hence $p_{X,Y,Z}(\cdot,\cdot,\cdot)$ remains unchanged;
\item the rate region remains the same;
\item
$$|\mathcal{V}'|\leq |\mathcal{X}|+1.$$
\end{itemize}
Let $\mathbb{P}$ be a connected compact subset of probability mass functions on $\mathcal{X}$. Let $\mathcal{X}=\{x_1,x_2,\cdots x_{|\mathcal{X}|}\}$ and consider the following continuous functions on $\mathbb{P}$:
\begin{align}\label{fctgj}
g_j(p_X(\cdot))  &= \left\{ \begin{array}{ll}
 p_X(x_j) &  j=1,\cdots,|\mathcal{X}|-1\\
H(X|W,Z)& j=|\mathcal{X}|\\
H(W|Z)& j=|\mathcal{X}|+1.
\end{array}
\right. 
\end{align}
The first $|\mathcal{X}|-1$ functions are trivially continuous with respect to $p_X(\cdot)$. Functions $$H(X|W,Z)= \sum \limits_{w,z}H(X|W=w,Z=z)p_{W,Z}(w,z) ,$$ and $$H(W|Z)= \sum \limits_{z}H(W|Z=z)p_Z(z)$$  are continuous with respect to $p_X(\cdot)$ due to the continuity of $p_{X|W,Z}(\cdot|\cdot,\cdot)$, $p_{W,Z}(\cdot,\cdot)$, $p_{W|Z}(\cdot|\cdot)$ and $p_Z(\cdot)$ with respect to $p_X(\cdot)$ since
$$p_{W,Z}(w,z)=\sum \limits_{x}p_{W,Z|X}(w,z|x)p_X(x),$$ 
$$p_{X|W,Z}(x|w,z)=p_{W,Z|X}(w,z|x)\frac{p_X(x)}{p_{W,Z}(w,z)},$$
$$p_Z(z)=\sum \limits_{x}p_{Z|X}(z|x)p_X(x),$$
and
$$p_{W|Z}(w|z)=\sum \limits_x p_{W|Z,X}(w|z,x)p_{Z|X}(z|x)\frac{p_X(x)}{p_Z(z)}.$$
Now, due to Support Lemma \cite[Appendix C, Page 631]{ElgaKim12}, there exists $V' \sim p_{V'}(\cdot)$ with $|\mathcal{V}'|\leq |\mathcal{X}|+1$ and a collection of conditional probability mass functions $p_{X|V'}(\cdot|v')\in \mathbb{P}$, indexed by $v' \in \mathcal{V}'$ such that for $j=1,\cdots, |\mathcal{X}|+1$,
$$\int_{\mathcal{V}}g_j(p_{X|V}(\cdot|v))dF(v)=\sum \limits_{v'\in \mathcal{V}'}g_j(p_{X|V'}(\cdot|v'))p_{V'}(v').$$
Hence, by \eqref{fctgj} we have
\begin{align}
\int_{\mathcal{V}}p_{X|V}(x_i|v)dF(v)&=p_X(x_i)=\sum \limits_{v'}p_{X|V'}(x_i|v')p_{V'}(v'), \label{eq:conservedPmf} \nonumber \\
&\hspace{15 ex}1 \leq i \leq |\mathcal{X}| \\
H(X|V,W,Z)&=\int_{\mathcal{V}}H(X|V=v,W,Z)dF(v) \nonumber \\
&=\sum \limits_{v'}H(X|V'=v',W,Z)p_{V'}(v')\nonumber \\
&=H(X|V',W,Z), \label{eq:conservedXVWZ}\\
H(W|V,Z)&=\int_{\mathcal{V}}H(W|V=v,Z)dF(v)\nonumber \\
&=\sum \limits_{v'}H(W|V'=v',Z)p_{V'}(v') \nonumber \\
&=H(W|V',Z). \label{eq:conservedWVZ} 
\end{align}
Moreover due to \eqref{eq:conservedPmf} and the Markov chain 
$$V-X-(Y,W,Z),$$
$p_X$ remains unchanged if we change $V$ to $V'$, hence the
joint probability $p_{X,Y,W,Z}(x,y,w,z)$ and the related quantities $H(X|W,Z)$, $H(W|Y)$, and $H(W|Z)$ are preserved. This implies that  the rate region obtained by changing  $V$ to $V'$ remains unchanged.
\end{IEEEproof}

\begin{IEEEproof} [Proof of Theorem \ref{th:OuterBound1}]

Since the rate pair $ (R_X,R_Y) $ is achievable, 
there exists a decoding function\footnote{$ \varphi_X(\mathbf{X}) $ and $ \varphi_Y(\mathbf{Y}) $ are received 
messages from transmitters.} 
$$ \psi(\varphi_X(\mathbf{X}),\varphi_Y(\mathbf{Y}),\mathbf{Z})=\mathbf{U} $$ 
such that 
\begin{align}
P(\mathbf{U} \neq f(\mathbf{X},\mathbf{Y},
\mathbf{Z})) \rightarrow 0  \text{ as } 
 n \rightarrow \infty. \label{eq:infGeneralOuterBound}
\end{align}

Define the distortion function 
\begin{gather}
\raisetag{-10pt}
d(x,y,z,u) = \begin{cases}
0 \mbox{  if } u=f(x,y,z),\\
1 \mbox{ otherwise. }  
\end{cases} \nonumber
\end{gather}

Since 
\begin{align*}
P(\mathbf{U} \neq f(\mathbf{X},\mathbf{Y},\mathbf{Z})) 
\geq P(U_i \neq f(X_i,Y_i,Z_i)),
\end{align*}  we have
\begin{align}
 P(\mathbf{U} \neq f(\mathbf{X},\mathbf{Y},\mathbf{Z})) 
 &\geq \frac{1}{n} \sum \limits_{i=1}^n p(U_i \neq f(X_i,Y_i,Z_i))\nonumber \\
 &\defeq d(\mathbf{X},\mathbf{Y},\mathbf{Z},\mathbf{U}). \label{eq:geqGeneralOuterBound}
\end{align}
From \eqref{eq:infGeneralOuterBound} and \eqref{eq:geqGeneralOuterBound}, $ d(\mathbf{X},\mathbf{Y},\mathbf{Z},\mathbf{U}) \rightarrow 0 $  as $ n\rightarrow \infty $.

Hence, assuming the same distortion for both sources, $ (R_X,R_Y) \in R_D(0,0), $\footnote{$ R_D(D_X,D_Y)$ is the rate distortion region 
for correlated source $ X $ and $ Y $ 
with distortion criteria $ D_X $ and $ D_Y $, respectively.}
 according to \cite[Theorem~5.1]{Tung78}
 there exist some random variables $ V $ and $ W $ 
 and a function $ g(V,W,Z) $ such that $$ \mathbb{E}d(X,Y,Z,g(V,W,Z)) = 0$$ 
  $$ V-X-(Y,Z)$$ $$ (X,Z)-Y-W, $$  
 and
 \begin{align}
 R_X &\geq I(X,Y;V|W,Z) \nonumber \\
 R_Y &\geq H(X,Y;W|V,Z) \nonumber \\ 
 R_X+R_Y &\geq I(X,Y;V,W|Z) \nonumber .
\end{align}

It remains to show that $ (V,W)$ satisfy $H(f(X,Y,Z)|V,W,Z)=0$. 

Since the distortion is equal to $ 0 $, for any 
$ (v,x_1,y_1,w,z) $ and $ (v,x_2,y_2,w,z) $ that satisfy
$$p(v,x_1,y_1,w,z)\cdot p(v,x_2,y_2,w,z) > 0,$$ 
we should have
$$ f(x_1,y_1,z)=g(v,w,z)=f(x_2,y_2,z). $$ 
This implies that $H(f(X,Y,Z)|V,W,Z)=0$ by  Lemma \ref{lem:jointdef} given in the Appendix.
\end{IEEEproof}

\begin{IEEEproof}[Proof of Corollary~\ref{th:OuterBound2}] 
To show the first inequality of the corollary, note that if $(R_X,R_Y)$ is an achievable rate pair for  $(X,Y,Z)$, then $(R_X,0)$ is an achievable rate pair for $(X,Constant,(Y,Z))$, {\it{i.e.}}, for the setting where $Y$ is revealed to the receiver.  The first inequality of Theorem~\ref{th:OuterBound1} for $(X,Constant,(Y,Z))$ becomes
\begin{align}
R_X &\geq I(X;V|W,Y,Z) \nonumber 
\end{align}
for some $V$ and $W$ that satisfy
\begin{align}
&H(f(X,Y,Z)|V,W,Y,Z)=0 \label{eq:Hcor1} \\
&\hspace{13mm} V - X - (Y,Z) \nonumber \\
&\hspace{3mm}(X,Y,Z) - Constant - W \label{eq:XYZConsW} \\
&\hspace{9mm}(V,W)-X-(Y,Z). \label{eq:VWXYZ}
\end{align}
Therefore,
\begin{align}
R_X &\geq I(X;V|W,Y,Z)\nonumber \\
&= H(X|W,Y,Z)-H(X|V,W,Y,Z)\nonumber \\
& \stackrel{(a)}{\geq} H(X|Y,Z)-H(X|V,W,Y,Z)\nonumber \\
&=I(X;V,W|Y,Z)\label{eq:firstTerm}
\end{align}
where $(a)$ holds due to \eqref{eq:XYZConsW}.

Moreover, \eqref{eq:Hcor1}, \eqref{eq:VWXYZ} and Lemma~\ref{lem:JointCompressionV2Y} gives
$$X \in (V,W) \in \Gamma(G_{X|Y,Z}).$$
This together with \eqref{eq:firstTerm} and Definition~\ref{def:condGraphEnt})
yields
$$R_X \geq H_{G_{X|Y,Z}}(X|Y,Z).$$

The second inequality of the corollary can be derived similarly. 

For the third inequality note that $H(f(X,Y,Z)|V,W,Z)=0$, hence the Markov chain $(V,W)-(X,Y)-Z$ and  Lemma~\ref{lem:JointCompressionV2Y} gives
$$(X,Y) \in (V,W) \in \Gamma(G_{X,Y|Z}).$$
This together with Definition~\ref{def:condGraphEnt} and the third inequality of Theorem~\ref{th:OuterBound1} gives the desired result.

\end{IEEEproof}

\begin{IEEEproof}[Proof of Theorem~\ref{cor:independent}] For the converse of Theorem~\ref{cor:independent}, note that the two
inequalities in the corollary correspond to the first two inequalities of
Corollary~\ref{th:OuterBound2}.  

For achievability, suppose $V$ and $W$ satisfy the conditions of
Theorem~\ref{achiev}, {\it{i.e.}}, \eqref{eq:achieveMarkov} and  \eqref{eq:achieveSet1} holds.
 From two Markov chains \eqref{eq:achieveMarkov} and the fact that $X$ and $Y$ are independent given
$Z$, we deduce the long Markov chain
$$V-X-Z-Y-W\,.$$
It then follows that $$I(V;X|W,Z)=I(V;X|Y,Z)$$
and 
$$I(Y;W|V,Z)=I(Y;W|X,Z)\,.$$
Using Theorem~\ref{achiev}, we deduce that the rate
pair $(R_X,R_Y)$ given by
$$R_X=I(V;X|Y,Z)$$
and
$$R_Y=I(Y;W|X,Z)$$
is achievable. Now, since $X$ and
$Y$ are independent given $Z$,
$G_{Y|V,Z} =G_{Y|X,Z}$ by Claim c. of Lemma~\ref{lem:samecondgraph}. This allows
to minimize the above two mutual information terms separately, which shows that
the rate pair
$$R_X=\min_{\substack{X\in V \in \text{M}(\Gamma(G_{X|Y,Z})) \\V-X-(Y,Z)}}I(V;X|Y,Z)$$
$$R_Y=\min_{\substack{Y\in W \in \text{M}(\Gamma(G_{Y|X,Z})) \\W-Y-(X,Z)}}I(Y;W|X,Z)$$
 is achievable (Notice that $I(V;X|Y,Z)$ is a function of the joint distribution
 $p(v,x,y,z)$ only, thus the minimization constraint $V-X-(Y,Z,W)$ reduces to
 $V-X-(Y,Z)$. A similar comment applies to the minimization of $I(Y;W|X,Z)$.) The result then follows from Definition~\ref{def:condGraphEnt}.
\end{IEEEproof}

To prove Theorem~\ref{CapacityOnecomplete} we need the following definition:
\begin{definition} [Support set of a random variable] \label{def:SupportSet}
Let $ (V,X) \sim p(v,x) $ where $V$ is a random variable taking values in some countable set $\mathcal{V}=\{v_1,v_2,\cdots\}$. Define the random variable $S_X(V)$ as $(J,S)$ where $J$ is a random variable taking values in the positive integers $\{1,2,\ldots\}$, where $S$ is a random variable that takes values over the subsets of $\cal{X}$, and such that $(J,S)=(j,s)$ if and only if $V=v_j$ and  
$s=\{x : p(v_j,x)>0\}$.
\end{definition}
Note that $V$ and $S_X(V)$ are in one-to-one correspondance by definition. In the sequel, with a slight abuse of notation we write 
$Z\in S_X(V)$ whenever $Z$ is a random variable that takes values over the subsets of $\cal{X}$ and such that $ Z=s$ whenever the second index of $S_X(V)$ is $s$---{\it{i.e.}}, whenever $S_X(V)=(j,s)$ for some $j\in \{1,2,\ldots\}$.

\begin{IEEEproof}[Proof of Theorem \ref{CapacityOnecomplete}]
The proof uses Theorem~\ref{achiev} and a the result of \cite{BergYeu89}. An alternative proof using the canonical theory developped in \cite{JanaBla08} is provided in the Appendix. 

For the achievablity part of the theorem it suffices to let $V=X$ in Theorem~\ref{achiev}. Now for the converse.

Since the rate pair $ (R_X,R_Y) $ 
is achievable, there exist a decoding function\footnote{$ \varphi_X(\mathbf{X}) $ and $ \varphi_Y(\mathbf{Y}) $ are received 
messages from transmitters.}  
$$ \psi( \varphi_X(\mathbf{X}), \varphi_Y(\mathbf{Y}),\mathbf{Z})=\mathbf{U}, $$ 
such that 
\begin{align}
P(\mathbf{U} \neq f(\mathbf{X},\mathbf{Y},\mathbf{Z})) \rightarrow 0  
\text{ as }  n \rightarrow \infty. \label{eq:pzero}
\end{align}
Also, since $ f$ is partially invertible with respect to $ X $, 
{\it{i.e}}, $ X $ is a function of $ f(X,Y,Z) $ and 
$ Z $, there exist a function 
$$ g(\varphi_X(\mathbf{X}), \varphi_Y(\mathbf{Y}),\mathbf{Z})=(\hat{X}_1,..,\hat{X}_n)=\hat{\mathbf{X}}, $$ 
such that 
$$ P(\hat{\mathbf{X}} \neq \mathbf{X}) \rightarrow 0 \text{ as }
 n \rightarrow \infty.$$
Define the distortion measures
\begin{gather}
\raisetag{-10pt}
d_{X}(x,\hat{x}) = \begin{cases}
0 \mbox{  if } x=\hat{x}\\
1  \mbox{ otherwise}  
\end{cases} \nonumber
\end{gather}
and
\begin{gather}
\raisetag{-10pt}
d_{Y}(x,y,z,u) = \begin{cases}
0 \mbox{  if } u=f(x,y,z)\\
1  \mbox{ otherwise.}  
\end{cases} \nonumber
\end{gather}

Since $$ P(\mathbf{U} \neq f(\mathbf{X},\mathbf{Y},\mathbf{Z})) 
\geq P(U_i \neq f(X_i,Y_i,Z_i)),$$ we have
\begin{align}
 P(\mathbf{U} \neq f(\mathbf{X},\mathbf{Y},\mathbf{Z})) 
 &\geq \frac{1}{n} \sum \limits_{i=1}^n p(U_i \neq f(X_i,Y_i,Z_i))\nonumber \\
 &\defeq d_Y(\mathbf{X},\mathbf{Y},\mathbf{Z},\mathbf{U}). \label{eq:pdistortion}
\end{align}
From \eqref{eq:pzero} and \eqref{eq:pdistortion}, $ d_Y(\mathbf{X},\mathbf{Y},\mathbf{Z},\mathbf{U}) \rightarrow 0 $  as $ n\rightarrow \infty $.
With the same argument one shows that
$$d_X(\mathbf{X},\hat{\mathbf{X}})\triangleq 
\frac{1}{n}\sum \limits_{i=1}^nd_X(X_i,\hat{X}_i) \rightarrow 0 
\text{ as }n\rightarrow \infty.  $$
According to  
 \cite[Theorem~1]{BergYeu89} and its immediate extension to the case where there is side information $Z$ at the receiver, it follows that 
 there exists a random variable $ W' $ and two
  functions $g_1(X,W',Z)$ and $g_2(X,W',Z)$ 
 such that\footnote{There is one caveat in applying the converse 
arguments of \cite[Theorem~1]{BergYeu89}.
In our case we need the distortion measures to be 
defined over functions of the sources. More precisely, we need  Hamming distortion for
source $X$ and Hamming distortion for a function defined over both sources $(X,Y)$ and side information $Z$. However, it is straightforward to 
extend the converse of \cite[Theorem~1]{BergYeu89} to handle this setting (same as \cite{Yama82} which shows that Wyner and Ziv's result
\cite{WyneZiv76} can be extended to the case where the distortion measure is defined over a function
of the source and the side information.).
}
  $$\mathbb{E}d_X(X,g_1(X,W',Z))=0$$ $$\mathbb{E}d_Y(X,Y,Z,g_2(X,W',Z))=0$$ 
 $$ (X,Z)-Y-W', $$ 
  and
 \begin{align}
 R_X &\geq H(X|W',Z) \nonumber \\
 R_Y &\geq I(Y;W'|X,Z) \nonumber \\
 R_X+R_Y &\geq H(X|Z)+ I(Y;W'|X,Z). \label{eq:BergerYeung} 
\end{align}
Notice that since the distortion $ \mathbb{E}d_Y(X,Y,Z,g_2(X,W',Z)) $ is equal to zero,
for any $(x,y_1,w',z) $ and $ (x,y_2,w',z) $  that satisfy
 $$ p(x,y_1,w',z)\cdot p(x,y_2,w',z) >0 $$   we should have
$$ f(x,y_1,z)=g_2(x,w',z)=f(x,y_2,z). $$ 
This, according to  Lemma \ref{lem:jointdef} 
 in the Appendix, is equivalent to $$H(f(X,Y,Z)|X,W',Z)=0.$$
  
Since $H(f(X,Y,Z)|X,W',Z)=0$ and since $ (X,Z)-Y-W' $ forms a Markov chain, 
using Corollary \ref{lem:JointCompressionV2Y} in the Appendix yields
$$ Y \in S_Y(W') \in \text{M}(\Gamma(G_{Y|X,Z})), $$
  and $$ (X,Z)-Y-S_Y(W'). $$
  
  Also, by definition of $S_Y(W') $ (Definition \ref{def:SupportSet}) we have
 \begin{align}
 H(X|W',Z)&=H(X|S_Y(W'),Z) \nonumber \\
 I(Y;W'|X,Z)&=I(Y;S_Y(W')|X,Z) \nonumber \\
 H(X|Z)+ I(Y;W'|X,Z)&=H(X|Z)+ I(Y;S_Y(W')|X,Z) \label{eq:BergerEqual}.
\end{align}
Taking $ W=S_Y(W') $ and using \eqref{eq:BergerYeung} 
and \eqref{eq:BergerEqual} completes the proof.

\noindent{\it{Remark:}} Note that due to Definition~\ref{def:SupportSet}, $S_Y(W')=(J,S)$ takes different values for $w'_1,w'_2 \in W'$ even if $\{y:p(w'_1,y)>0\}=\{y:p(w'_2,y)>0\}$, \textit{i.e.}, $w'_1$ and $w'_2$ with the same $S$ index but different $J$ indices. This is unlike the converse for the point-to-point case \cite[Proof of Theorem~2]{OrliRoc01}, where such a $w'_1$ and $w'_2$ are considered as one sample $S$. By considering them as one sample we always have 
\begin{align*}
I(Y;W'|X,Z) &= I(Y;S_Y(W')|X,Z)\\
&=I(Y;J,S|X,Z) \\& 
\geq I(Y;S|X,Z),
\end{align*}
 but we have 
\begin{align*}
H(X|W',Z) &= H(X|S_Y(W'),Z)\\&
=H(X|J,S,Z)\\
&\leq H(X|S,Z)
\end{align*} which means that $R_Y \geq I(Y;S|X,Z) $ holds but $R_X \geq H(X|S,Z)$ may not hold. This is why the reduction to sets (and so to maximal independent sets) are possible in point-to-point case but it may not be possible for correlated sources case.
\end{IEEEproof}

\small{
\bibliographystyle{plain}
\bibliography{Bibliography}}

\appendix
\begin{lemma} \label{lem:jointdef}
Let $$ (V,X,Y,W,Z) \in  \mathcal{V } \times \mathcal{X} \times \mathcal{Y} 
\times \mathcal{W} \times \mathcal{Z}$$ be distributed 
according to $ p(v,x,y,w,z) $. The two following
statements are equivalent:
\begin{itemize}
\item[a)]\ $H(f(X,Y,Z)|V,W,Z)=0.$
\item[b)] For all $$ (x_1,y_1,z), (x_2,y_2,z) \in \mathcal{X} \times \mathcal{Y} \times \mathcal{Z} $$ 
$$ (v,w) \in \mathcal{V} \times \mathcal{W} $$ 
that satisfy
$$ p(v,x_1,y_1,w,z) \cdot p(v,x_2,y_2,w,z) > 0, $$ 
we have $$ f(x_1,y_1,z) = f(x_2,y_2,z). $$
\end{itemize} 
\end{lemma}

\begin{IEEEproof}
For showing the equivalence notice 
that $ H(f(X,Y,Z)|V,W,Z)=0 $ if and only 
if there exist a function $ g(v,w,z) $ such that
$$ f(X,Y,Z)=g(V,W,Z),$$
which is the same as b.
\end{IEEEproof}
\begin{lemma} \label{lem:jointCompressionLong}
Given $ (X,Y,Z) \sim p(x,y,z) $ and $ f(X,Y,Z) $, random variables 
$ (V,W) $ satisfy 
\begin{align}
H(f(X,Y,Z)|V,W,Z)=0 \nonumber
\end{align}
 and the Markov chains
\begin{align}
V-X-(Y,W,Z)  \nonumber \\
 (V,X,Z)-Y-W, \label{eq:jointCompressMarkov1}
\end{align}
$$  $$ 
if and only if they satisfy\footnote{See Definition~\ref{def:SupportSet} p.~\pageref{def:SupportSet} for the definition of the support set of a random variable.} 
\begin{align}
&X \in S_X(V)\in \text{M}(\Gamma(G_{X|Y,Z})) \nonumber \\
Y &\in S_Y(W)\in \text{M}(\Gamma(G_{Y|S_X(V),Z})), \nonumber
\end{align}
 and 
 \begin{align}
 S_X(V)-X-(Y,S_Y(W),Z) \nonumber \\
(S_X(V),X,Z)-Y-S_Y(W). \label{eq:jointCompressMarkov2}
 \end{align}
 \end{lemma}
By letting $ V=X $ in Lemma~\ref{lem:jointCompressionLong} 
we get the following Corollary:
\begin{corollary} \label{lem:JointCompressionV2Y}
Given $ (X,Y,Z) \sim p(x,y,z) $ and $ f(X,Y,Z) $, 
$ W $ satisfies $$H(f(X,Y,Z)|X,W,Z)=0,$$ 
and $$ (X,Z)-Y-W,$$  
if and only if  $$ Y \in S_Y(W) \in  \text{M}(\Gamma(G_{Y|X,Z})),$$ 
and  $$ (X,Z)-Y-S_Y(W). $$
\end{corollary}

\begin{IEEEproof} [Proof of Lemma \ref{lem:jointCompressionLong}]
The lemma is a direct consequence of the following four claims, proved thereafter:
\begin{itemize}
\item[a.] $ X \in S_X(V) $ and $ Y \in S_Y(W) $ always hold.
\item[b.] Markov chians \eqref{eq:jointCompressMarkov1} and \eqref{eq:jointCompressMarkov2} are equivalent.
\end{itemize}
Further, when
 these  Markov chains hold, Claims c. and d. below hold:
\begin{itemize}
\item[c.] $ (V,W) $ satisfy 
$$H(f(X,Y,Z)|V,W,Z)=0,$$ 
if and only if for all $x_1,x_2 \in S_X(v) $ and
$y_1,y_2 \in S_Y(w) $ such that  
 $$ p(x_1,y_1,z)\cdot p(x_2,y_2,z)>0,$$ it holds that
$$ f(x_1,y_1,z)=f(x_2,y_2,z). $$

\item[d.] $$S_X(V)\in \text{M}(\Gamma(G_{X|Y,Z})) $$
$$  S_Y(W)\in \text{M}(\Gamma(G_{Y|S_X(V),Z})), $$
if and only if for all $x_1,x_2 \in S_X(v) $ and
$y_1,y_2 \in S_Y(w) $ that
 $$ p(x_1,y_1,z)\cdot p(x_2,y_2,z)>0,$$  it holds that 
$$ f(x_1,y_1,z)=f(x_2,y_2,z). $$
\end{itemize}

Claims a. and b. are direct consequences of Definition \ref{def:SupportSet}. We now establish Claims c. and d..
\begin{itemize}
\item[c.] Notice that due to the Markov chains \eqref{eq:jointCompressMarkov1}
we can write $$ p(v,x,y,w,z)=p(x,y,z)\cdot p(v|x)\cdot p(w|y). $$
Hence $$ p(v,x_1,y_1,w,z)\cdot p(v,x_2,y_2,w,z) > 0, $$ 
if and only if $$ p(x_1,y_1,z)\cdot p(x_2,y_2,z) \cdot p(v,x_1) \cdot p(v,x_2)
\cdot p(y_1,w)\cdot p(y_2,w)>0  ,$$ which is equivalent to the conditions 
$$ x_1,x_2 \in S_X(v) $$ $$ y_1,y_2 \in S_Y(w) $$ 
$$ p(x_1,y_1,z) \cdot p(x_2,y_2,z)>0.$$

Using the Lemma \ref{lem:jointdef} completes the proof of the claim.
\item[d.]
The proof for the converse part follows from Definition \ref{def:GenCondGraph}.

To prove the direct part, for
$ x_1,x_2 \in S_X(v) $, $ y_1,y_2 \in S_Y(w) $ such that
$$ p(x_1,y_1,z)\cdot p(x_2,y_2,z)>0, $$ we show that 
$$ f(x_1,y_1,z)=f(x_2,y_2,z).$$ 
\begin{itemize}
\item[i.] If $ y_1=y_2 $, then since 
$$ S_X(v) \in  \text{M}(\Gamma(G_{X|Y,Z})), $$ 
for $ x_1,x_2 \in S_X(v) $, 
$ f(x_1,y_1,z) = f(x_2,y_2,z)$ (the same argument 
is valid if$ x_1=x_2 $.).
\item[ii.] If $ x_1\neq x_2 $, 
$ y_1 \neq y_2$, then from 
$$ S_Y(w) \in  \text{M}(\Gamma(G_{Y|S_X(V),Z})) $$ we have
\begin{align*}
f(x_1,y_1,z) &= \tilde{f}_X(S_X(v),y_1,z)\\
&= \tilde{f}_X(S_X(v),y_2,z)=f(x_2,y_2,z).
\end{align*} 
\end{itemize}
\end{itemize}
 \end{IEEEproof}

We present a proof that establishes Theorem~\ref{CapacityOnecomplete} using the canonical theory developed in
\cite{JanaBla08}.  For the
cardinality bound $|{\mathcal{W}}|\leq |\mathcal{Y}|+1$ one should repeat the same argument as the one given at the end of the proof of Theorem~\ref{achiev}. Suppose there is a third transmitter who knows $ U=f(X,Y,Z) $ and
sends some information with rate $ R_U $  to the receiver. For this problem, the rate
region is the set of achievable rate pairs $ (R_X,R_Y,R_U) $. By intersecting this rate
region with $ R_U=0 $, we obtain the rate region for our two transmitter computation
problem.

Consider the three transmitter setting as above. Since $f(X,Y,Z) $ is partially
invertible, we can equivalently assume that the goal for the receiver is to 
obtain $ (X,U)$. This corresponds  to $(M,J,L)=(3,2,0)$ in the
Jana-Blahut notation, and, using \cite[Theorem 6]{JanaBla08}, the rate region is given by
the set of all $(R_X,R_Y,R_U)$ such that
\begin{align}
R_X & \geq H(X|W',Z,U) \nonumber \\
R_Y & \geq I(Y;W'|X,Z,U) \nonumber \\
R_U & \geq H(U|X,W',Z) \label{eq:ruzerro} \\
R_X+R_Y & \geq I(X,Y;X,W'|Z,U) \nonumber \\
R_X+R_U & \geq I(X,U;X,U|W',Z)\nonumber \\
&=H(X|W',Z)+H(U|X,W',Z) \nonumber \\
R_Y+R_U & \geq I(Y,U;W',U|X,Z)\nonumber \\
&=I(Y;W'|X,Z)+I(U;W'|X,Y,Z)\nonumber \\ &\hspace{2ex}+H(U|X,W',Z) \nonumber \\
R_X+R_Y+R_U & \geq I(X,Y,U;X,W',U|Z) \nonumber \\
&=I(X,Y,U;X,W'|Z)+I(X,Y,U;U|X,W',Z) \nonumber \\
&=I(X,Y;X,W'|Z)+I(U;X,W'|X,Y,Z)\nonumber \\
 &\hspace{2ex} +H(U|X,W',Z), \label{eq:rate3}
\end{align}
for some $ W' $ that satisfies $$ (X,Z,U)-Y-W' .$$  Due to this Markov chain we have
\begin{align}
I(U;W'|X,Y,Z)= I(U;X,W'|X,Y,Z)=0 . \label{eq:dataprocess}
\end{align} 
Intersecting with $ R_U=0 $, from \eqref{eq:ruzerro}
we derive that 
\begin{align}
H(U|X,W',Z)=0. \label{eq:intersect}
\end{align}  
Hence, using \eqref{eq:dataprocess} and \eqref{eq:intersect}, the last three inequalities in \eqref{eq:rate3}
become
\begin{align}
R_X+0 & \geq H(X|W',Z)\geq H(X|W',Z,U) \nonumber \\
R_Y+0 & \geq I(Y;W'|X,Z) \nonumber \\
&=H(W'|X,Z)-H(W'|X,Y,Z) \nonumber \\
&=H(W'|X,Z)-H(W'|X,Y,Z,U) \nonumber \\
&\geq H(W'|X,Z,U)-H(W'|X,Y,Z,U)\nonumber \\
&=I(Y;W'|X,Z,U) \nonumber 
\end{align}
\begin{align}
R_X+R_Y+0 & \geq I(X,Y;X,W'|Z)=H(X|Z)+I(Y;W'|X,Z) \nonumber \\
&\geq H(X|Z,U)+I(Y;W'|X,Z,U)\nonumber \\
&=I(X,Y;X,W'|Z,U)\,,\nonumber
\end{align} 
which also imply the first three inequalities in \eqref{eq:rate3}.

Therefore, when the three last inequalities of \eqref{eq:rate3} hold and when $ 
H(U|X,W',Z)=0 $, all other inequalities are satisfied. The rate region for the two
transmitter problem thus becomes the set
of rate pairs $ (R_X,R_Y) $ that satisfy
\begin{align}
R_X & \geq H(X|W',Z) \nonumber \\
R_Y & \geq I(Y;W'|X,Z)  \nonumber \\
R_X+R_Y & \geq I(X,Y;X,W'|Z), \nonumber
\end{align} 
 for some $ W' $ that satisfies $ (X,Z)-Y-W' $ and $  H(U|X,W',Z)=0 $. Now, according to
 Corollary~\ref{lem:JointCompressionV2Y}, we have $$ Y \in S_Y(W') \in  \text{M}(\Gamma(G_{Y|X,Z})),$$ 
and  $$ (X,Z)-Y-S_Y(W'').$$ Taking $ W=S_Y(W') $ completes the proof. 
\end{document}